\documentclass[prb,superscriptaddress,twocolumn,longbibliography,floatfix]{revtex4-1}

\usepackage{graphicx}
\usepackage{bm}
\usepackage{braket}
\usepackage{hyperref}
\usepackage[usenames]{xcolor}
\usepackage{amsmath}
\usepackage{graphicx}
\usepackage{latexsym}
\usepackage{amsmath,amssymb}
\usepackage{xcolor}
\usepackage{stackrel}
\usepackage{accents}
\usepackage{empheq}
\usepackage[title]{appendix}
\usepackage{comment}

\usepackage{float}
\usepackage{braket}
\usepackage{bbold}
\usepackage{mathtools}

\begin{document}

\title{Unsupervised machine learning for detecting mutual independence among eigenstate regimes in interacting quasiperiodic chains}
\author{Colin Beveridge}
\affiliation{Department of Physics, University of Notre Dame, South Bend, IN 46556 USA}
\author{Kathleen Hart}
\affiliation{Department of Physics, University of Notre Dame, South Bend, IN 46556 USA}
\author{Cassio Rodrigo Cristani}
\affiliation{Department of Mathematics and Physics,
	 Catholic University of the Sacred Heart, Brescia, Italy}
\author{Xiao Li}
\affiliation{Department of Physics, City University of Hong Kong, Kowloon, HongKong, China}
\author{Enrico Barbierato}
\affiliation{Department of Mathematics and Physics,
	 Catholic University of the Sacred Heart, Brescia, Italy}
\email{enrico.barbierato@unicatt.it}
\author{Yi-Ting Hsu}
\affiliation{Department of Physics, University of Notre Dame, South Bend, IN 46556 USA}
\email{yhsu2@nd.edu}
\date{\today}
\begin{abstract}
Manybody eigenstates that are neither thermal nor manybody-localized (MBL) were numerically found in certain interacting chains with moderate quasiperiodic potentials. The energy regime consisting of these non-ergodic but extended (NEE) eigenstates has been extensively studied for being a possible manybody mobility edge between the energy-resolved MBL and thermal phases. Recently, the NEE regime was proposed to be a prethermal phenomenon that generally occurs when different operators spread at sizably different timescales. Here, we numerically examine the mutual independence among the NEE, MBL, and thermal regimes in the lens of eigenstate entanglement spectra (ES).  
Given the complexity and rich information embedded in ES, we develop an unsupervised learning approach that is designed to quantify the mutual independence among general phases. Our method is first demonstrated on an illustrative toy example that uses RGB color data to represent phases, then applied to the ES of an interacting generalized Aubry Andre model from weak to strong potential strength. We find that while the MBL and thermal regimes are mutually independent, the NEE regime is dependent on the former two and smoothly appears as the potential strength decreases. We attribute our numerically finding to the fact that the ES data in the NEE regime exhibits both an MBL-like fast decay and a thermal-like long tail.
\end{abstract}

\maketitle
\textit{Introduction---}
Eigenstates of an interacting one-dimensional (1D) chain with sufficiently weak random disorders are known to become thermalized even in the absence of a heat bath \cite{heatbath1,heatbath2,heatbath3}. 
As the disorder strength $W$ increases, the system crossovers into a wide prethermal regime \cite{PrethermalRandom_Chandran} before transitioning into the manybody localized (MBL) phase \cite{AvalancheRandom_Huse}.  
This prethermal regime under an intermediate $W$ still thermalizes (although after a longer thermalization time) \cite{PreTH_longTHtime_PRE1,PreTH_longTHtime_PRE2,PreTH_longTHtime_PRX,PreTH_longTHtime_PRL1,PreTH_longTHtime_PRL2}, but simultaneously exhibits non-thermal behaviors, such as Poisson-like level statistics \cite{AvalancheRandom_Huse}.  
In addition to randomly disordered chains, interacting chains with deterministic quasiperiodic potentials were also numerically\cite{AA1,GAAU_PRL2015,intGPD_MBL_EElevelreturntransport,GAAU_PRB2016,AA2,intGAA_nonGPD,ML_GAA_Hsu,butterflyAA,longrangeAA,intt1t2,Tu_AA_avalanche} and experimentally\cite{AAexp1,AAexp2,GAAexp1} found to exhibit thermal and MBL phases. 
Moreover, at certain moderate potential strength $\lambda$, 
numerical evidences have revealed an energy regime between the energy-resolved thermal and MBL phases where the manybody eigenstates are non-ergodic but extended. 
This non-thermal and non-MBL regime was dubbed a non-ergodic extended regime (NEE), and was first observed in 
a generalized Aubry Andre model called Ganeshan-Pixley-Das Sarma (GPD) model \cite{GPD} in the presence of interaction \cite{GAAU_PRL2015,GAAU_PRB2016}. 
More generally, the NEE regime was recently proposed to be a generic prethermal phenomenon in finite-size systems that occurs when different opertators spread at sizably different  timescales\cite{Tu_NEE_prethermal}. The identification of the NEE regime is thus crucial for further understandings of the prethermal physics in finite-size systems  accessible by experimental and numerical studies.
\begin{figure}[t!]
{\includegraphics[width=0.45\textwidth]{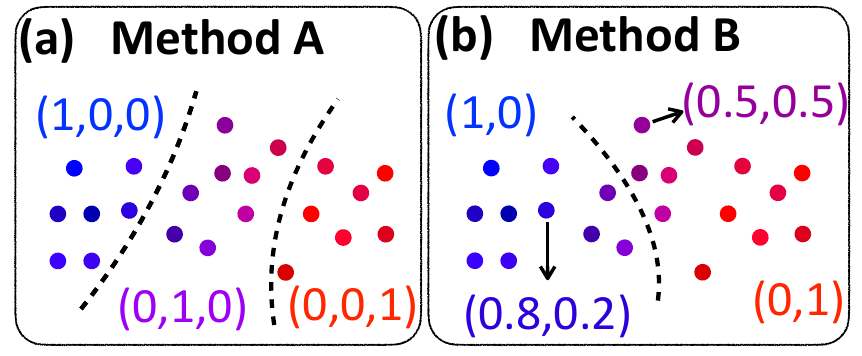}}
\caption{Illustration for the difference between the two unsupervised clustering methods. The same set of input data represented by color dots are subjected to (a) K-means clustering, and (b) Fuzzy clustering, which forbid and allow fractional memberships, respectively. In this schematic scenario, Method A and B find three and two clusters, respectively. Each data point $j$ is labeled by a  membership vector $c_{j,k}$ for cluster $k$. In (a), the membership vector has $k=3$ components and $c_{j,k}=0,1$. The purple data points forms a third cluster besides the blue and red clusters. In (b), the membership vector has $k=2$ components and $0<c_{j,k}<1$. The purple data points are distributed into the blue or red cluster, depending on different degrees of fractional memberships.}
\label{Fig2}
\end{figure}

Due to the lack of well-defined order parameters for NEE, its range (in energy $E$ and $\lambda$) has been loosely identified as a non-thermal and non-MBL regime based on inconsistency among various inequivalent diagnostics, such as entanglement entropy \cite{GAAU_PRL2015,intGPD_MBL_EElevelreturntransport,GAAU_PRB2016,intGAA_nonGPD,NEM_avalanche_Tu}, dynamics\cite{intGPD_MBL_EElevelreturntransport,intGAA_nonGPD,NEM_avalanche_Tu,Tu_NEE_prethermal}, and fluctuations in observables within a small window \cite{GAAU_PRL2015,GAAU_PRB2016}.
However, it remains puzzling how to precisely determine the number of eigenstate `phases' (defined in finite-size systems) between thermal and MBL phases using multiple diagnostics. This is because different diagnostics generally exhibit different MBL-to-thermal transition energies, creating multiple energy regimes that do not behave consistently among all diagnostics. Therefore, some of us proposed in Ref. \onlinecite{ML_GAA_Hsu} to determined the number of eigenstate phases 
by identifying the number of patterns in a single informationally rich quantity, the entanglement spectra (ES) of eigenstates. 
Given the complexity of ES, the determination was performed using a \textit{`self-supervised' machine learning (ML) approach}, which does not require prior knowledge about the number of ES patterns. 
This approach found three different regimes labeled by ES patterns in a GPD model with a weak to moderate potential strength $\lambda$, suggesting a phase diagram of energy-resolved MBL, NEE, and thermal phases. 

How independent the NEE regime is of the more well-defined thermal and MBL phases, however, remains elusive. 
While prior works have shown that NEE exhibits a third type of distinct behavior that is neither MBL nor thermal \cite{ML_GAA_Hsu,NEM_avalanche_Tu}, such behavior could also be a consequence of averaging over a mixture of MBL-like or thermal-like eigenstates. 


In this work, we propose a novel numerical approach that determines the number of eigenstate phases and characterizes how independent these phases are by combining two \textit{unsupervised} learning algorithms \cite{lloyd1982least,bezdek_pattern_1981}: 
\begin{itemize}
    \item Method A: K-Means clustering
    \item Method B: Fuzzy clustering. 
\end{itemize}  
As shown schematically in Fig. 1, both methods self-consistently distribute the input data into groups by minimizing the ‘intra-group distance’ and maximizing the ‘inter-group distance’.
However, while Method A \cite{lloyd1982least} assigns each data point to a definite cluster, Method B \cite{bezdek_pattern_1981} allows \textit{fractional memberships}.
When applied together to determine a phase diagram of interest, Method A determines the qualitatively different regimes in some parameter space along with their phase boundaries, whereas Method B further determines if a phase is independent or dependent of the other phases. See Supplementary Materials (SM) section III for a brief review for these two well-established numerical methods. 

In the following, as a benchmark, we will first examine the energy-resolved phase diagram of 
an interacting GPD model under different potential strengths $\lambda$ using the well-developed self-supervised ML method \cite{ML_GAA_Hsu} (see Fig. 2a and b). 
Next, before investigating the GPD model using the unsupervised clustering approach, we will demonstrate our approach on an illustrative toy example that uses RBG color vectors to represent phases (see Fig. 3). In particular, in the blue-purple-red case, we find that Method A sees three clusters but Method B sees only two clusters since purple is a mixture of blue and red. 
Finally, we will apply our unsupervised approach to the interacting GPD model to examine the phase diagram.  
With ES being the input datasets, we find that each ES from the NEE regime exhibits features from both the MBL and thermal regimes. 
Benchmarking with the self-supervised phase diagram , our unsupervised results provide information about the nature of the NEE regime in an interacting quasiperiodic system. 

\textit{Model---} 
Our goal is to understand the manybody eigenstate phases in quasiperiodic chains with single-particle mobility edges, under a short-range density-density interaction. As a case study, we consider a generalized Aubry-Andre model $H=H_0+H_{\text{int}}$ called the GPD model, where the kinetic and interaction terms are given by  
\begin{align}
& H_0 = \sum_{j=1}^L \left(-t(c_{j}^\dag c_{j+1} + H.c.) + 2\lambda \frac{\cos(2\pi qj + \phi)}{1 - \alpha \cos(2\pi qj + \phi)} n_{j} \right)\nonumber\\
& H_{int} = V\sum_{j=1}^{L} n_{j+1} n_{j},  
\end{align}
respectively. Here, $n_{j}=c_{j}^{\dag} c_j$ is the fermionic density operator at site $j$,
$t$ is the nearest-neighbor hopping strength, and $V$ is the nearest-neighbor interaction strength. The quasiperiodic potential described by the second term in $H_0$ has a strength of 2$\lambda$, an irrational wave number $q = 2/(1+\sqrt{5})$, and a chosen global phase $\phi\in[0,2\pi)$. In the rest of this article, we express all energies in the unit of $t$ and set the dimensionless parameter $\alpha =-0.8$ to obtain a substantial energy range for the NEE regime. Moreover, 
the ES of the manybody eigenstates of $H$
are obtained using Lanzcos method at a system size $L=30$ with $1/6$ filling.
Specifically, we cut the full spectrum into 125 bins, evenly distributed between the maximum and minimum eigenvalues of the given spectrum. We then sample 5 eigenstates per bin per $\phi$, where $\phi$ takes 40 equally spaced values between 0 and $2\pi$. We choose not to include all eigenstates in the datasets since a full diagonalization is memory prohibitive.  
At an intermediate potential strength $\lambda=0.3\equiv\lambda^*$, it was reported in Ref. \onlinecite{ML_GAA_Hsu} that the interacting manybody spectrum found by the self-superpervised ML method contains three energy regimes in which the eigenstates show different ES patterns (see Fig. 2b). 

\begin{figure}[t!]
{\includegraphics[width=0.5\textwidth]{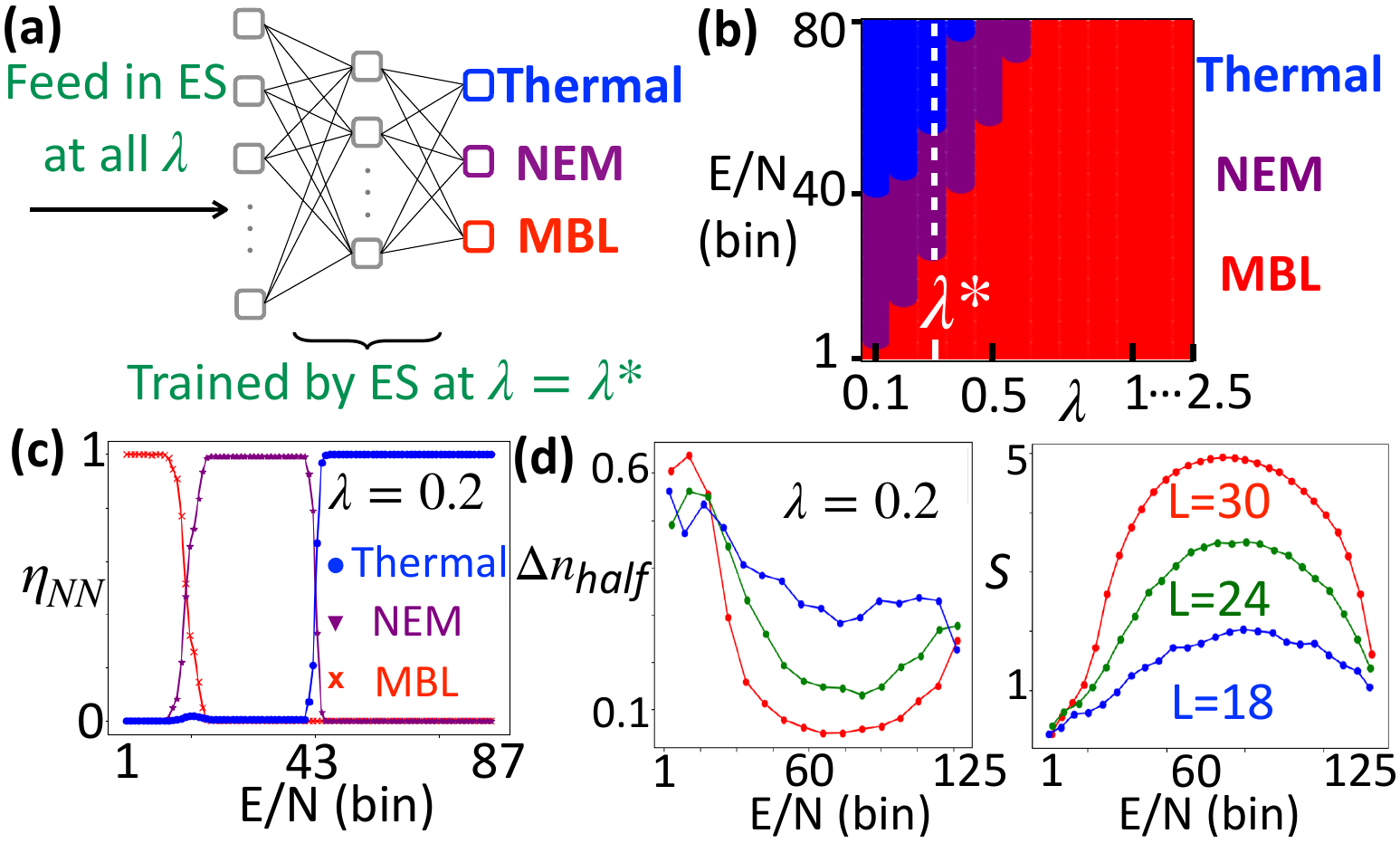}}
\caption{(a) The schematics and (b) the resulting GPD model phase diagram of the self-supervised network approach \cite{ML_GAA_Hsu}. The phases at potential strength $\lambda^*=0.3$ along the white dashed line is consistent with that reported in Ref. \cite{ML_GAA_Hsu}. (c) The confidence level $\eta_{NN}\in [0,1]$ of the neural network in (a) at potential strength $\lambda=0.2$, where substantial energy regimes for all three types of eigenstates are identified with nearly perfect confidence $\eta_{NN}\sim 1$. (d) The variance in the half-chain density $\Delta n_{\text{half}}$ and the entanglement entropy $S$ as a function of energy per electron $E/N$ at different system sizes $L$. The calculated $\eta_{NN}$, $\Delta n_{\text{half}}$, and $S$ for other potential strengths $\lambda$ are shown in SM section I.}
\label{Fig1}
\end{figure}

\textit{Self-supervised phase diagram---} 
As a benchmark for the unsupervised approach, we now apply the well-developed self-supervised ML method \cite{ML_GAA_Hsu} to examine how  the three eigenstate phases at $\lambda=\lambda^*$ evolve with the potential strength $\lambda$. 
First, we obtain the three-phase network classifier that distinguishes the MBL, thermal, and NEE ES patterns at $\lambda=\lambda^*$ with a nearly perfect confidence level by reproducing the phase diagram reported in Ref. \onlinecite{ML_GAA_Hsu} (along the white dashed line at $\lambda=\lambda^*$ in Fig. 2b). Such a three-phase classifier is a feed-forward artificial neural network, where the input layer takes the ES of a manybody eigenstate and the output layer contains three neurons. The three output neurons characterize the probabilities $p_i$ of the input ES being in phase $i=$ MBL, thermal, NEE, with the normalization condition  $\sum_i p_i=1$ (see Fig. 2a).   
Note that without assuming there are three phases at $\lambda=\lambda^*$, we have checked but find low confidence levels for the two-phase and four-phase scenarios, where there are only MBL and thermal phases, and where there is an additional fourth phase, respectively.
We therefore stick to the three-phase classifier trained at $\lambda=\lambda^*$. By feeding in the ES data from the full manybody spectra at different potential strengths $\lambda\neq\lambda^*$, we obtain the phase diagram in Fig. 2b. 
Physically speaking, this phase diagram is generated by asking the network classifier how well a given ES from energy $E$ at a general potential strength $\lambda\neq \lambda^*$ resembles the ES from the MBL, NEE, and thermal regimes at $\lambda=\lambda^*$. 

In Fig. 2b, for a wide potential strength around $\lambda^{*}$, 
the low, intermediate, and mid-spectra \footnote{Eigenstates of the top band edge of the spectrum resemble those from the bottom band edge of the spectrum.} regimes exhibit MBL, NEE, and thermal-like ES behaviors, respectively. 
As $\lambda$ decreases from $\lambda^*$, the three-phase scenario smoothly transitions towards a two-phase scenario that consists of predominantly NEE and thermal regimes. We attribute the tiny MBL-like regime on the spectrum edge at $\lambda=0.1$ to the finite size effect. 
In contrast, as $\lambda$ increases from $\lambda^*$, the three-phase scenario smoothly transitions into a two-phase scenario with only NEE and MBL regimes, then eventually into a one-phase scenario with only the MBL regime. 

Note that instead of existing only right at $\lambda^*$, where the network classifier is trained, NEE extends to a finite range in $\lambda$ with a nearly perfect confidence (Fig. 2c) and smoothly disappears. 
This shows that the NEE regime at $\lambda^*$ is unlikely an artifact from the network convergence at $\lambda^*$.
We further verify in Fig. 2d the extended and non-ergodic properties of the NEE regime at $\lambda\neq\lambda^*$ using the energy-resolved entanglement entropy $S(E)$ and the variance in half-chain density $n_{\text{half}}(E)$, respectively, consistent with previous findings \cite{GAAU_PRL2015}. 
Specifically, we find that the eigenstates in the NEE regime exhibit a volume-law $S(E)$ and large variances of $n_{\text{half}}(E)$ within small energy windows. These two properties show that these metallic eigenstates are also non-ergodic.
What remains elusive from this self-supervised ML study is  whether the NEE regime  can be viewed as a wide transition regime between MBL and thermal phases, due to the nature of the transition or finite-size effects, or  as a distinct phase with ES features fully independent of the other two. 


\textit{Unsupervised clustering approach---}
We now introduce our clustering approach using an illustrative toy example, where we take the RGB color codes as the input datasets. Specifically, the RGB color code represents each color using a vector ($r$,$g$,$b$), where $r$, $g$, $b$ are real numbers between 0 and 1. We construct two datasets Set I and Set II: Set I consists of three groups of linearly independent colors, red, green, and blue, represented by $(r_1, 0,0)$, $(0, g_1,0)$, and $(0,0,b_1)$, respectively. The data are generated by assigning $1-\epsilon < r_1, g_1, b_1 <1$ with a slightly varying small real number $\epsilon$ so that the data points visually appear as red, green, or blue with tiny variations. Set II consists of red, blue, and a wide range of purple, represented respectively by $(r_2,0,0)$, $(0,0,b_2)$, and $(\tilde{r}_2,0,\tilde{b}_2)$. The red and blue data points are generated in the same way as in Set I: $1-\epsilon < r_2, b_2 <1$, whereas the purple data points are generated by assigning $\tilde{r}_2$ and $\tilde{b}_2$ to be random values  ranged from 0 to 1 (see SM section II).  The two sets of colors are visually shown as color bars in Fig. 3a and b.

\begin{figure}[t!]
{\includegraphics[width=0.4\textwidth]{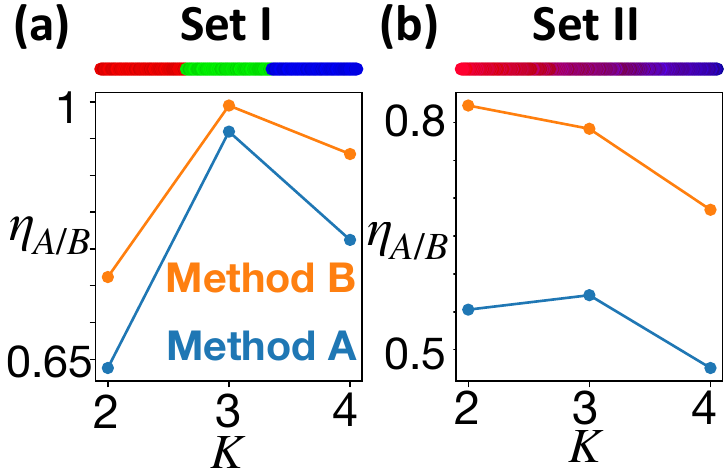}}
\caption{Clustering results for the toy example that uses RGB color vectors as the input datasets. The confidence indicators $\eta_{A/B}$ obtained by Method A and B are shown for datasets (a) Set I and (b) Set II. The corresponding colors of the data points in each dataset are shown in the color bar. The indicators $\eta_{A/B}$ are obtained under the assumption of different cluster numbers $K=2,3,4$. The $K$ with the largest $\eta_{A (B)}$ is the cluster number that Method A (B) deems most probable.}
\label{Fig1}
\end{figure}

Next, we employ the two clustering methods Method A\cite{lloyd1982least} and Method B\cite{bezdek_pattern_1981} to determine the number of colors (`phases') in datasets Set I and Set II, and explain how the clustering results can identify the difference in nature of Set I and Set II. 
Clustering is an unsupervised learning method that separates input data into different clusters according to the “distance” among data points. 
Conventionally, the cluster number $K$ is treated as a chosen hyperparameter, along with the initialization of cluster centers and the convergence tolerance. 
However, in our case, the cluster number $K$ is the unknown information that we want to identify.  
To this end, we first perform one clustering procedure for each possible $K$, then identify the most probable $K$ as the cluster number with the highest ‘confidence indicator' $\eta (K)$ for the convergence of the procedure. The precise definitions of the confidence indicators $\eta_{A/B} (K)$ for Method A and B are given in SM section V.

The task is to calculate the confidence indicators $\eta_{A/B}(K)$ using Method A and B for the two datasets Set I and II. 
Method A is the Silhouette method in K-Means clustering\cite{lloyd1982least}. 
Given an assumed $K$, K-Means algorithm distributes all data points into $K$ distinct clusters by iteratively updating cluster centers to simultaneously minimize and  maximize the intra- and inter-cluster sum of squared distances, respectively.  
We then correspondingly update the integer membership degree $c_{j,k}=0,1$ of each data point $j$ to each cluster $k=1,\cdots,K$.
We repeat this procedure  assuming different cluster numbers, $K=2,3,4$, 
where the initial cluster centers are chosen randomly. 
At the end of each procedure, $c_{j,k}$ for all $j$ and $k$ will converge to either 0 or 1, and the convergence level will be  quantified using  Silhouette coefficient $\eta_A (K) \in [-1,1]$. We then determine the most probable cluster number as the $K$ that maximizes the average silhouette coefficient $\eta_A (K)$.

Method B is C-Fuzzy clustering \cite{bezdek_pattern_1981}, which is a K-Means-like algorithm that further accommodates noisy datasets, where data points might not exclusively belong to a single cluster. 
In practice, C-Fuzzy clustering assigns each data point $j$ with a fractional membership degree across multiple clusters.  
In contrast to Method A, the membership degree $c_{j,k}$ is now a real number between 0 and 1, calculated by a membership function that depends on the distance between data point $j$ and each cluster center $k$. 
The more a data point belongs to a cluster, the more influence it has on determining the cluster center in the convergence process. 
The algorithm iterates the steps of assigning membership degrees and updating the cluster centers until the cluster centers converge.   
Similar to Method A, in Method B we repeatedly perform the procedure assuming different cluster numbers $K=2,3,4$. The convergence level of each procedure is labeled by a confidence indicator called fuzzy partition coefficient (FPC) index $\eta_B (K)$. 
The most probable cluster number is then determined by the $K$ that maximizes $\eta_B (K)$.

The convergence indicators $\eta_\text{A} (K)$ and $\eta_\text{B} (K)$ for datasets Set I and II are shown in Fig. 3.  
For the red-green-blue dataset Set I, both $\eta_\text{A} (K)$ and $\eta_\text{B} (K)$ peak at $K=3$ (see Fig. 3a). 
This shows that the clustering algorithm robustly finds three independent clusters regardless of whether a fractional membership is allowed, as expected for three groups of colors represented by orthogonal vectors.  
In sharp contrast, for the red-purple-blue dataset Set II, although $\eta_\text{A} (K)$ still peaks at $K=3$, $\eta_\text{B} (K)$ now peaks at $K=2$ (see Fig. 3b). 

This discrepancy can be intuitively understood as follows. 
When a fractional membership is allowed, the purple points, which are  linear combinations of blue and red, do not form a distinct cluster independent of red and blus clusters.  
We therefore propose that for a given dataset where the cluster number is unknown, by applying both unsupervised clustering algorithms Method A and Method B, one can determine (1) the number of distinct clusters of data, and (2) how many of which are independent clusters. Specifically, the discrepancy between Method A and B identifies 'mixed' clusters containing data points that emulate features from multiple other clusters. 
Such a mixed regime, which carries drastically fluctuating properties that emulate traits from multiple other regimes, could indicate a wide phase transition or a crossover in a physics system.


\begin{figure}[t!]
{\includegraphics[width=0.5\textwidth]{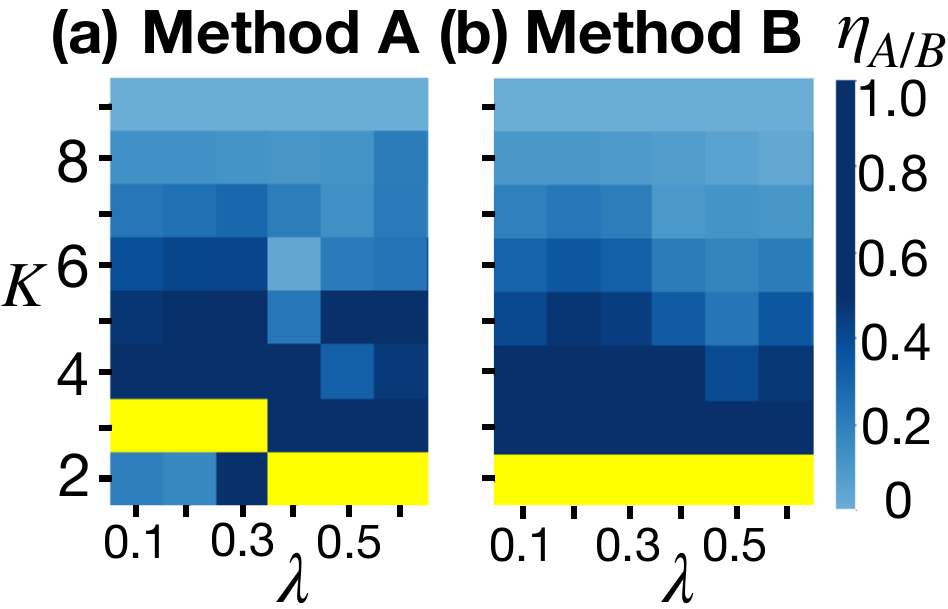}}
\caption{Clustering results for the GPD model. The confidence indicators $\eta_{A/B}(K,\lambda)$ obtained by (a) Method A and (b) Method B are shown in color maps for different possible cluster numbers $K$ and  potential strengths $\lambda$. At each $\lambda$, the most probable cluster number $K_{A/B}(\lambda)$ is identified by the largest indicator $\eta_{A/B}$, which is colored in yellow. Explicit values of $\eta_{A/B}(K,\lambda)$ are shown in SM section IV.}
\label{Fig1}
\end{figure}

\textit{Non-ergodic metal regime---} 
We now perform clustering Method A and B to determine the number of eigenstate phases in interacting GPD model in Eq. 1, benchmarking the self-supervised phase diagram we found in Fig. 2b. In particular, by comparing the number of phases found by the two methods, we can inspect how independent the NEE regime is from  the MBL and thermal phases. 

To construct the input datasets, we employ Lanszos method to diagonalize an $L=30$ chain, filled with $N=5$ electrons. 
For each potential strength $\lambda$, we generate the ES of manybody eigenstates sampled throughout the full spectrum $E$ and sort each ES in descending order.  
 Although each ES contains 4944 components, we extract the portion that are informationally rich to minimize the numerical noise and to improve the clustering performance.  
 Specifically, we first omit the nearly zero elements within numerical noise range with a cutoff at $10^{-10}$, then perform a dimensional reduction analysis on the remaining 250 ES components. For simplicity, we perform Principal Component Analysis (PCA)\cite{pearson1901liii} and keep the two largest PCA eigen-modes. We have checked that the two largest eigen-modes are responsible for at least 80$\%$ of the data variation for most of the $\lambda$ values so that essential information is likely preserved (see SM section III for a brief introduction to PCA and our data).  

Method A is performed on this ES datasets to determine the most probable number of clusters $K_A(\lambda)$ at each potential strength $\lambda$ in the following way.  
First, we systematically perform one K-Means clustering procedure for each $K=2,3,\cdots,9$. The cluster number $K$ corresponds to the number of ES patterns, which also indicates the number of eigenstate phases. 
For each clustering procedure, we quantify the convergence level using the Silhouette coefficients $\eta_\text{A}(K,\lambda)\in [0,1]$. 
At each $\lambda$, the most probable cluster number $K_A(\lambda)$ is determined by the $K$ with the largest $\eta_A$ (labeled in yellow in Fig. 4a). 
We find that $K_A(\lambda)$ in Fig. 4a is consistent with the number of eigenstate phases found in our self-supervised phase diagram in Fig. 2b at every potential strength $\lambda$ smaller than $0.7$, where there are more than one energy-resolved phases. For datasets containing only one cluster, clustering methods are generally not applicable since they tend to show low confidence for all $K$ {\cite{tibshirani2001estimating}}. The consistency between our self-supervised and unsupervised results confirms that there are indeed three ES patterns at weak to moderate potential strengths $\lambda$. Nonetheless, the independence among these three ES patterns is yet to be examined by Method B. 

Next, we perform Method B on the same input ES dataset to determine the number of clusters $K_B(\lambda)$ at each potential strength $\lambda$, where fractional membership is allowed. 
Similar to Method A, we perform one Fuzzy-C clustering procedure for each possible cluster number $K=2,3,\cdots,9$, where the corresponding convergence level is quantified by the FPC index  $\eta_B(K,\lambda)\in[0,1]$. 
The most probable cluster number  $K_B(\lambda)$ is then identified by the highest $\eta_B$ at each $\lambda$. 
In Fig. 4b, we show the confidence indicator $\eta_B(K,\lambda)$, where the most probable number of clusters $K_B(\lambda)$ is highlighted in yellow. 
Although $K_B(\lambda)$ agrees well with $K_A(\lambda)$ at stronger potential strengths $\lambda$, they unexpectedly differ at weak to moderate $\lambda$’s. 
In contrast to the phase diagrams generated by the self-supervised learning and unsupervised Method A, we find that \textit{there is no $\lambda$ that exhibits three eigenstate phases} in the lens of Method B with fractional membership. Instead, Method B finds only two phases in that potential range, i.e. $K_B(\lambda)=2$ whenever $K_A(\lambda)=3$ (compare Fig. 4a and b). 

\begin{figure}[t!]
{\includegraphics[width=0.45\textwidth]{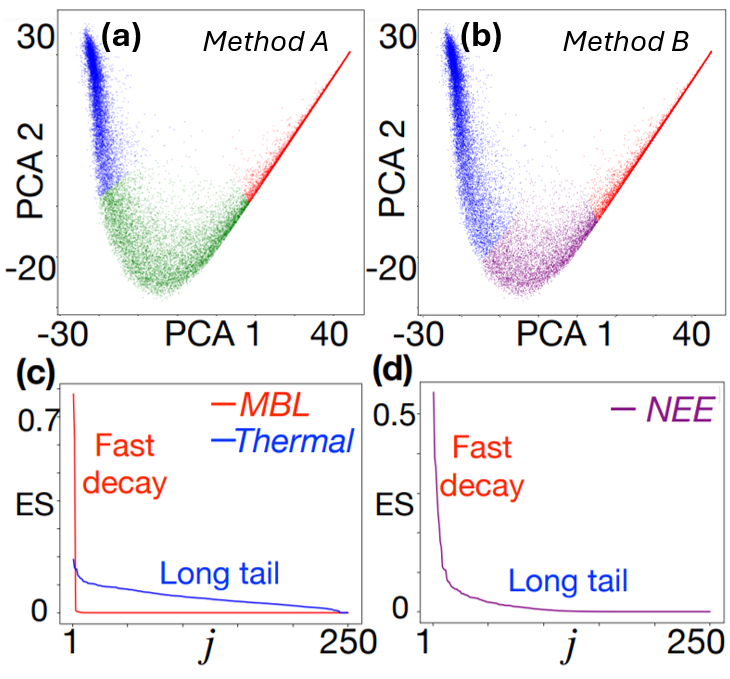}}
\caption{(a)(b) The ES data of the interacting GPD model at $\lambda=0.3$, constructed from the eigenstates throughout the spectrum, are displayed in the two-dimensional space spanned by the first two components in the PCA analysis. In (a), each data point is colored  according to which of the three clusters found by Method A it belongs to. In (b), the data points with full memberships in the two clusters Method B found are colored in blue and red. The data points with fractional memberships are colored in purple. ES data for example data points chosen from the (c) red, blue, and (d) purple regimes in (b). In (c), the ES data of red and blue points  feature a fast decay at small component $i$ and a long tail at large $i$, respectively, which resemble the features of MBL and thermal eigenstates. In (d), the ES data of a purple point concurrently show both features.}
\label{Fig5}
\end{figure}

This discrepancy could be intuitively understood by the fractional membership that is allowed only in Method B but not A.  
If a data point $i$ has non-zero fractional membership $0<c_{i,k}<1$ for multiple clusters $k$, it implies that the corresponding ES exhibits features from more than one eigenstate phase.
Therefore, when Method B sees fewer clusters than Method A, it implies that the clusters obtained from A are not all independent of each other. 
To verify this interpretation from the actual data, 
we now focus on the inconsistent regime $\lambda\in[0.1,0.3]$ where Method A finds three phases but Method B finds two phases. 
Specifically, we examine the distribution of ES data in Method A and B by representing each ES with a point in the two-dimensional space spanned by the first two PCA eigen-components (see Fig. 5a and b). 

In Fig. 5a with $K_A=3$ clusters, we color a data point $i$ in red, blue, or green if it carries a membership vector $\textbf{c}_i=(c_{i,1}, c_{i,2}, c_{i,3})=$ (1,0,0), (0,0,1), or (0,1,0), respectively, where $c_{i,k}=0, 1$ labels the membership in cluster $k$.   
We identify the red cluster $k=1$ as the MBL regime since the ES data therein (see Fig. 5c) are consistent with the typical fast power-law decay in an MBL ES \cite{ES_MBL}. 
Similarly, we identify the blue cluster $k=2$ as the thermal regime since the ES data therein have the typical Marchenko-Pastur-like form expected from a thermalized state \cite{ES_ETH}. 
Therefore, the ES in the green cluster $k=3$ are not MBL- nor thermal-like in the lens of Method A, and we attribute the green cluster to the NEE regime.

In Fig. 5b with $K_B=2$ clusters, we color data point $i$ in red and blue if it carries a full membership $\textbf{c}_i = (c_{i,1}, c_{i,2})=(1-\epsilon,\epsilon)$ and $(\epsilon,1-\epsilon)$, respectively, with a threshold $\epsilon<0.2$. 
For the rest of the points with a fractional membership $\textbf{c}_i$ with $\epsilon\geq0.2$, we color them in purple. 
Note that the data distribution here in Fig. 5b nearly align with that in Fig. 5a when the green is replaced by the purple, besides points near the cluter boundaries.  
This shows that the two independent clusters that Method B identifies are the MBL (red) and thermal (blue) regimes. 
Moreover, the green NEE regime found by Method A mainly consists of and contains all the ES data points with fractional memberships, which implies that the ES data in the NEE regime exhibit certain MBL-like and thermal-like features simultaneously. 
Specifically, we find that the ES data with fractional membership all exhibit a faster decay at small component i than the Marchenko-Pastur distribution from the thermal-like ES. Moreover, at large component $i$, they all exhibit a longer tail than the MBL-like ES. 
Therefore, we attribute the fast decay and the long tail in ES to be the MBL-like and thermal-like features that Method B recognizes, respectively. 
These two features coexist in all the ES data points in the purple regime at different degrees. 

\textit{Conclusion and discussion ---} Our unsupervised clustering method provides a general numerical approach for identifying the mutual independence among nearby phases. In this work, we demonstrate our method on a finite-length GPD chain by identifying the dependence of the NEE regime on energy-resolved MBL and thermal phases. 
We choose ES as the ML input data since ES contains nearly complete information about the eigenstates without requiring much human pre-processing or assumptions. Moreover, ES is also  computationally affordable when using the Lanczos algorithm for diagonalizing the Hamiltonian. These two advantages make ES an ideal input for unbiased ML analyses. Interesting future directions include applying our unsupervised and self-supervised ML approaches to investigate the independence among many-body eigenstate phases in other quasiperiodic or disordered systems with a single-particle mobility edge, such as quasiperiodic models with longer range hopping\cite{t1t2}, as well as comparing the ML results to other well-known diagnostics, such as multifractal dimensions. 
Beyond this case study, our method provides a general numerical tool for studying the prethermal phenomena in non-equilibrium systems with either random disorders or quaiperiodic potentials based on ES and other eigenstate properties.

\textit{Acknowledgement---} Y.-T.H. thanks Yi-Ting Tu and Dinhduy Vu for very helpful discussions. Y.-T.H. acknowledges support from NSF Grant No. DMR-2238748. 
This work was performed in part at Aspen Center for Physics, which is supported by National Science Foundation grant PHY-2210452. This work was supported in part by the National Science Foundation under Grant No. NSF PHY-1748958. X.L. is supported by the Research Grants Council of Hong Kong (Grants No. CityU 21304720, No. CityU 308 11300421, No. CityU 11304823, and No. C7012-21G), and City University of Hong Kong (Project No. 9610428).


\begin{thebibliography}{35}%
\makeatletter
\providecommand \@ifxundefined [1]{%
 \@ifx{#1\undefined}
}%
\providecommand \@ifnum [1]{%
 \ifnum #1\expandafter \@firstoftwo
 \else \expandafter \@secondoftwo
 \fi
}%
\providecommand \@ifx [1]{%
 \ifx #1\expandafter \@firstoftwo
 \else \expandafter \@secondoftwo
 \fi
}%
\providecommand \natexlab [1]{#1}%
\providecommand \enquote  [1]{``#1''}%
\providecommand \bibnamefont  [1]{#1}%
\providecommand \bibfnamefont [1]{#1}%
\providecommand \citenamefont [1]{#1}%
\providecommand \href@noop [0]{\@secondoftwo}%
\providecommand \href [0]{\begingroup \@sanitize@url \@href}%
\providecommand \@href[1]{\@@startlink{#1}\@@href}%
\providecommand \@@href[1]{\endgroup#1\@@endlink}%
\providecommand \@sanitize@url [0]{\catcode `\\12\catcode `\$12\catcode
  `\&12\catcode `\#12\catcode `\^12\catcode `\_12\catcode `\%12\relax}%
\providecommand \@@startlink[1]{}%
\providecommand \@@endlink[0]{}%
\providecommand \url  [0]{\begingroup\@sanitize@url \@url }%
\providecommand \@url [1]{\endgroup\@href {#1}{\urlprefix }}%
\providecommand \urlprefix  [0]{URL }%
\providecommand \Eprint [0]{\href }%
\providecommand \doibase [0]{http://dx.doi.org/}%
\providecommand \selectlanguage [0]{\@gobble}%
\providecommand \bibinfo  [0]{\@secondoftwo}%
\providecommand \bibfield  [0]{\@secondoftwo}%
\providecommand \translation [1]{[#1]}%
\providecommand \BibitemOpen [0]{}%
\providecommand \bibitemStop [0]{}%
\providecommand \bibitemNoStop [0]{.\EOS\space}%
\providecommand \EOS [0]{\spacefactor3000\relax}%
\providecommand \BibitemShut  [1]{\csname bibitem#1\endcsname}%
\let\auto@bib@innerbib\@empty
\bibitem [{\citenamefont {Deutsch}(1991)}]{heatbath1}%
  \BibitemOpen
  \bibfield  {author} {\bibinfo {author} {\bibfnamefont {J.~M.}\ \bibnamefont
  {Deutsch}},\ }\bibfield  {title} {\enquote {\bibinfo {title} {Quantum
  statistical mechanics in a closed system},}\ }\href {\doibase
  10.1103/PhysRevA.43.2046} {\bibfield  {journal} {\bibinfo  {journal} {Phys.
  Rev. A}\ }\textbf {\bibinfo {volume} {43}},\ \bibinfo {pages} {2046--2049}
  (\bibinfo {year} {1991})}\BibitemShut {NoStop}%
\bibitem [{\citenamefont {Srednicki}(1994)}]{heatbath2}%
  \BibitemOpen
  \bibfield  {author} {\bibinfo {author} {\bibfnamefont {Mark}\ \bibnamefont
  {Srednicki}},\ }\bibfield  {title} {\enquote {\bibinfo {title} {Chaos and
  quantum thermalization},}\ }\href {\doibase 10.1103/PhysRevE.50.888}
  {\bibfield  {journal} {\bibinfo  {journal} {Phys. Rev. E}\ }\textbf {\bibinfo
  {volume} {50}},\ \bibinfo {pages} {888--901} (\bibinfo {year}
  {1994})}\BibitemShut {NoStop}%
\bibitem [{\citenamefont {Rigol}\ \emph {et~al.}(2008)\citenamefont {Rigol},
  \citenamefont {Dunjko},\ and\ \citenamefont {Olshanii}}]{heatbath3}%
  \BibitemOpen
  \bibfield  {author} {\bibinfo {author} {\bibfnamefont {Marcos}\ \bibnamefont
  {Rigol}}, \bibinfo {author} {\bibfnamefont {Vanja}\ \bibnamefont {Dunjko}}, \
  and\ \bibinfo {author} {\bibfnamefont {Maxim}\ \bibnamefont {Olshanii}},\
  }\bibfield  {title} {\enquote {\bibinfo {title} {Thermalization and its
  mechanism for generic isolated quantum systems},}\ }\href {\doibase
  10.1038/nature06838} {\bibfield  {journal} {\bibinfo  {journal} {Nature}\
  }\textbf {\bibinfo {volume} {452}},\ \bibinfo {pages} {854--858} (\bibinfo
  {year} {2008})}\BibitemShut {NoStop}%
\bibitem [{\citenamefont {Long}\ \emph {et~al.}(2023)\citenamefont {Long},
  \citenamefont {Crowley}, \citenamefont {Khemani},\ and\ \citenamefont
  {Chandran}}]{PrethermalRandom_Chandran}%
  \BibitemOpen
  \bibfield  {author} {\bibinfo {author} {\bibfnamefont {David~M.}\
  \bibnamefont {Long}}, \bibinfo {author} {\bibfnamefont {Philip J.~D.}\
  \bibnamefont {Crowley}}, \bibinfo {author} {\bibfnamefont {Vedika}\
  \bibnamefont {Khemani}}, \ and\ \bibinfo {author} {\bibfnamefont {Anushya}\
  \bibnamefont {Chandran}},\ }\bibfield  {title} {\enquote {\bibinfo {title}
  {Phenomenology of the prethermal many-body localized regime},}\ }\href
  {\doibase 10.1103/PhysRevLett.131.106301} {\bibfield  {journal} {\bibinfo
  {journal} {Phys. Rev. Lett.}\ }\textbf {\bibinfo {volume} {131}},\ \bibinfo
  {pages} {106301} (\bibinfo {year} {2023})}\BibitemShut {NoStop}%
\bibitem [{\citenamefont {Morningstar}\ \emph {et~al.}(2022)\citenamefont
  {Morningstar}, \citenamefont {Colmenarez}, \citenamefont {Khemani},
  \citenamefont {Luitz},\ and\ \citenamefont {Huse}}]{AvalancheRandom_Huse}%
  \BibitemOpen
  \bibfield  {author} {\bibinfo {author} {\bibfnamefont {Alan}\ \bibnamefont
  {Morningstar}}, \bibinfo {author} {\bibfnamefont {Luis}\ \bibnamefont
  {Colmenarez}}, \bibinfo {author} {\bibfnamefont {Vedika}\ \bibnamefont
  {Khemani}}, \bibinfo {author} {\bibfnamefont {David~J.}\ \bibnamefont
  {Luitz}}, \ and\ \bibinfo {author} {\bibfnamefont {David~A.}\ \bibnamefont
  {Huse}},\ }\bibfield  {title} {\enquote {\bibinfo {title} {Avalanches and
  many-body resonances in many-body localized systems},}\ }\href {\doibase
  10.1103/PhysRevB.105.174205} {\bibfield  {journal} {\bibinfo  {journal}
  {Phys. Rev. B}\ }\textbf {\bibinfo {volume} {105}},\ \bibinfo {pages}
  {174205} (\bibinfo {year} {2022})}\BibitemShut {NoStop}%
\bibitem [{\citenamefont {\ifmmode~\check{S}\else \v{S}\fi{}untajs}\ \emph
  {et~al.}(2020)\citenamefont {\ifmmode~\check{S}\else \v{S}\fi{}untajs},
  \citenamefont {Bon\ifmmode~\check{c}\else \v{c}\fi{}a}, \citenamefont
  {Prosen},\ and\ \citenamefont {Vidmar}}]{PreTH_longTHtime_PRE1}%
  \BibitemOpen
  \bibfield  {author} {\bibinfo {author} {\bibfnamefont {Jan}\ \bibnamefont
  {\ifmmode~\check{S}\else \v{S}\fi{}untajs}}, \bibinfo {author} {\bibfnamefont
  {Janez}\ \bibnamefont {Bon\ifmmode~\check{c}\else \v{c}\fi{}a}}, \bibinfo
  {author} {\bibfnamefont {Toma\ifmmode \check{z}\else~\v{z}\fi{}}\
  \bibnamefont {Prosen}}, \ and\ \bibinfo {author} {\bibfnamefont {Lev}\
  \bibnamefont {Vidmar}},\ }\bibfield  {title} {\enquote {\bibinfo {title}
  {Quantum chaos challenges many-body localization},}\ }\href {\doibase
  10.1103/PhysRevE.102.062144} {\bibfield  {journal} {\bibinfo  {journal}
  {Phys. Rev. E}\ }\textbf {\bibinfo {volume} {102}},\ \bibinfo {pages}
  {062144} (\bibinfo {year} {2020})}\BibitemShut {NoStop}%
\bibitem [{\citenamefont {Sels}\ and\ \citenamefont
  {Polkovnikov}(2021)}]{PreTH_longTHtime_PRE2}%
  \BibitemOpen
  \bibfield  {author} {\bibinfo {author} {\bibfnamefont {Dries}\ \bibnamefont
  {Sels}}\ and\ \bibinfo {author} {\bibfnamefont {Anatoli}\ \bibnamefont
  {Polkovnikov}},\ }\bibfield  {title} {\enquote {\bibinfo {title} {Dynamical
  obstruction to localization in a disordered spin chain},}\ }\href {\doibase
  10.1103/PhysRevE.104.054105} {\bibfield  {journal} {\bibinfo  {journal}
  {Phys. Rev. E}\ }\textbf {\bibinfo {volume} {104}},\ \bibinfo {pages}
  {054105} (\bibinfo {year} {2021})}\BibitemShut {NoStop}%
\bibitem [{\citenamefont {Sels}\ and\ \citenamefont
  {Polkovnikov}(2023)}]{PreTH_longTHtime_PRX}%
  \BibitemOpen
  \bibfield  {author} {\bibinfo {author} {\bibfnamefont {Dries}\ \bibnamefont
  {Sels}}\ and\ \bibinfo {author} {\bibfnamefont {Anatoli}\ \bibnamefont
  {Polkovnikov}},\ }\bibfield  {title} {\enquote {\bibinfo {title}
  {Thermalization of dilute impurities in one-dimensional spin chains},}\
  }\href {\doibase 10.1103/PhysRevX.13.011041} {\bibfield  {journal} {\bibinfo
  {journal} {Phys. Rev. X}\ }\textbf {\bibinfo {volume} {13}},\ \bibinfo
  {pages} {011041} (\bibinfo {year} {2023})}\BibitemShut {NoStop}%
\bibitem [{\citenamefont {Sierant}\ \emph
  {et~al.}(2020{\natexlab{a}})\citenamefont {Sierant}, \citenamefont
  {Delande},\ and\ \citenamefont {Zakrzewski}}]{PreTH_longTHtime_PRL1}%
  \BibitemOpen
  \bibfield  {author} {\bibinfo {author} {\bibfnamefont {Piotr}\ \bibnamefont
  {Sierant}}, \bibinfo {author} {\bibfnamefont {Dominique}\ \bibnamefont
  {Delande}}, \ and\ \bibinfo {author} {\bibfnamefont {Jakub}\ \bibnamefont
  {Zakrzewski}},\ }\bibfield  {title} {\enquote {\bibinfo {title} {Thouless
  time analysis of anderson and many-body localization transitions},}\ }\href
  {\doibase 10.1103/PhysRevLett.124.186601} {\bibfield  {journal} {\bibinfo
  {journal} {Phys. Rev. Lett.}\ }\textbf {\bibinfo {volume} {124}},\ \bibinfo
  {pages} {186601} (\bibinfo {year} {2020}{\natexlab{a}})}\BibitemShut
  {NoStop}%
\bibitem [{\citenamefont {Sierant}\ \emph
  {et~al.}(2020{\natexlab{b}})\citenamefont {Sierant}, \citenamefont
  {Delande},\ and\ \citenamefont {Zakrzewski}}]{PreTH_longTHtime_PRL2}%
  \BibitemOpen
  \bibfield  {author} {\bibinfo {author} {\bibfnamefont {Piotr}\ \bibnamefont
  {Sierant}}, \bibinfo {author} {\bibfnamefont {Dominique}\ \bibnamefont
  {Delande}}, \ and\ \bibinfo {author} {\bibfnamefont {Jakub}\ \bibnamefont
  {Zakrzewski}},\ }\bibfield  {title} {\enquote {\bibinfo {title} {Thouless
  time analysis of anderson and many-body localization transitions},}\ }\href
  {\doibase 10.1103/PhysRevLett.124.186601} {\bibfield  {journal} {\bibinfo
  {journal} {Phys. Rev. Lett.}\ }\textbf {\bibinfo {volume} {124}},\ \bibinfo
  {pages} {186601} (\bibinfo {year} {2020}{\natexlab{b}})}\BibitemShut
  {NoStop}%
\bibitem [{\citenamefont {Iyer}\ \emph {et~al.}(2013)\citenamefont {Iyer},
  \citenamefont {Oganesyan}, \citenamefont {Refael},\ and\ \citenamefont
  {Huse}}]{AA1}%
  \BibitemOpen
  \bibfield  {author} {\bibinfo {author} {\bibfnamefont {Shankar}\ \bibnamefont
  {Iyer}}, \bibinfo {author} {\bibfnamefont {Vadim}\ \bibnamefont {Oganesyan}},
  \bibinfo {author} {\bibfnamefont {Gil}\ \bibnamefont {Refael}}, \ and\
  \bibinfo {author} {\bibfnamefont {David~A.}\ \bibnamefont {Huse}},\
  }\bibfield  {title} {\enquote {\bibinfo {title} {Many-body localization in a
  quasiperiodic system},}\ }\href {\doibase 10.1103/PhysRevB.87.134202}
  {\bibfield  {journal} {\bibinfo  {journal} {Phys. Rev. B}\ }\textbf {\bibinfo
  {volume} {87}},\ \bibinfo {pages} {134202} (\bibinfo {year}
  {2013})}\BibitemShut {NoStop}%
\bibitem [{\citenamefont {Li}\ \emph {et~al.}(2015)\citenamefont {Li},
  \citenamefont {Ganeshan}, \citenamefont {Pixley},\ and\ \citenamefont
  {Das~Sarma}}]{GAAU_PRL2015}%
  \BibitemOpen
  \bibfield  {author} {\bibinfo {author} {\bibfnamefont {Xiaopeng}\
  \bibnamefont {Li}}, \bibinfo {author} {\bibfnamefont {Sriram}\ \bibnamefont
  {Ganeshan}}, \bibinfo {author} {\bibfnamefont {J.~H.}\ \bibnamefont
  {Pixley}}, \ and\ \bibinfo {author} {\bibfnamefont {S.}~\bibnamefont
  {Das~Sarma}},\ }\bibfield  {title} {\enquote {\bibinfo {title} {Many-body
  localization and quantum nonergodicity in a model with a single-particle
  mobility edge},}\ }\href {\doibase 10.1103/PhysRevLett.115.186601} {\bibfield
   {journal} {\bibinfo  {journal} {Phys. Rev. Lett.}\ }\textbf {\bibinfo
  {volume} {115}},\ \bibinfo {pages} {186601} (\bibinfo {year}
  {2015})}\BibitemShut {NoStop}%
\bibitem [{\citenamefont {Modak}\ and\ \citenamefont
  {Mukerjee}(2015)}]{intGPD_MBL_EElevelreturntransport}%
  \BibitemOpen
  \bibfield  {author} {\bibinfo {author} {\bibfnamefont {Ranjan}\ \bibnamefont
  {Modak}}\ and\ \bibinfo {author} {\bibfnamefont {Subroto}\ \bibnamefont
  {Mukerjee}},\ }\bibfield  {title} {\enquote {\bibinfo {title} {Many-body
  localization in the presence of a single-particle mobility edge},}\ }\href
  {\doibase 10.1103/PhysRevLett.115.230401} {\bibfield  {journal} {\bibinfo
  {journal} {Phys. Rev. Lett.}\ }\textbf {\bibinfo {volume} {115}},\ \bibinfo
  {pages} {230401} (\bibinfo {year} {2015})}\BibitemShut {NoStop}%
\bibitem [{\citenamefont {Li}\ \emph {et~al.}(2016)\citenamefont {Li},
  \citenamefont {Pixley}, \citenamefont {Deng}, \citenamefont {Ganeshan},\ and\
  \citenamefont {Das~Sarma}}]{GAAU_PRB2016}%
  \BibitemOpen
  \bibfield  {author} {\bibinfo {author} {\bibfnamefont {Xiaopeng}\
  \bibnamefont {Li}}, \bibinfo {author} {\bibfnamefont {J.~H.}\ \bibnamefont
  {Pixley}}, \bibinfo {author} {\bibfnamefont {Dong-Ling}\ \bibnamefont
  {Deng}}, \bibinfo {author} {\bibfnamefont {Sriram}\ \bibnamefont {Ganeshan}},
  \ and\ \bibinfo {author} {\bibfnamefont {S.}~\bibnamefont {Das~Sarma}},\
  }\bibfield  {title} {\enquote {\bibinfo {title} {Quantum nonergodicity and
  fermion localization in a system with a single-particle mobility edge},}\
  }\href {\doibase 10.1103/PhysRevB.93.184204} {\bibfield  {journal} {\bibinfo
  {journal} {Phys. Rev. B}\ }\textbf {\bibinfo {volume} {93}},\ \bibinfo
  {pages} {184204} (\bibinfo {year} {2016})}\BibitemShut {NoStop}%
\bibitem [{\citenamefont {Khemani}\ \emph {et~al.}(2017)\citenamefont
  {Khemani}, \citenamefont {Sheng},\ and\ \citenamefont {Huse}}]{AA2}%
  \BibitemOpen
  \bibfield  {author} {\bibinfo {author} {\bibfnamefont {Vedika}\ \bibnamefont
  {Khemani}}, \bibinfo {author} {\bibfnamefont {D.~N.}\ \bibnamefont {Sheng}},
  \ and\ \bibinfo {author} {\bibfnamefont {David~A.}\ \bibnamefont {Huse}},\
  }\bibfield  {title} {\enquote {\bibinfo {title} {Two universality classes for
  the many-body localization transition},}\ }\href {\doibase
  10.1103/PhysRevLett.119.075702} {\bibfield  {journal} {\bibinfo  {journal}
  {Phys. Rev. Lett.}\ }\textbf {\bibinfo {volume} {119}},\ \bibinfo {pages}
  {075702} (\bibinfo {year} {2017})}\BibitemShut {NoStop}%
\bibitem [{\citenamefont {Nag}\ and\ \citenamefont
  {Garg}(2017)}]{intGAA_nonGPD}%
  \BibitemOpen
  \bibfield  {author} {\bibinfo {author} {\bibfnamefont {Sabyasachi}\
  \bibnamefont {Nag}}\ and\ \bibinfo {author} {\bibfnamefont {Arti}\
  \bibnamefont {Garg}},\ }\bibfield  {title} {\enquote {\bibinfo {title}
  {Many-body mobility edges in a one-dimensional system of interacting
  fermions},}\ }\href {\doibase 10.1103/PhysRevB.96.060203} {\bibfield
  {journal} {\bibinfo  {journal} {Phys. Rev. B}\ }\textbf {\bibinfo {volume}
  {96}},\ \bibinfo {pages} {060203} (\bibinfo {year} {2017})}\BibitemShut
  {NoStop}%
\bibitem [{\citenamefont {Hsu}\ \emph {et~al.}(2018)\citenamefont {Hsu},
  \citenamefont {Li}, \citenamefont {Deng},\ and\ \citenamefont
  {Das~Sarma}}]{ML_GAA_Hsu}%
  \BibitemOpen
  \bibfield  {author} {\bibinfo {author} {\bibfnamefont {Yi-Ting}\ \bibnamefont
  {Hsu}}, \bibinfo {author} {\bibfnamefont {Xiao}\ \bibnamefont {Li}}, \bibinfo
  {author} {\bibfnamefont {Dong-Ling}\ \bibnamefont {Deng}}, \ and\ \bibinfo
  {author} {\bibfnamefont {S.}~\bibnamefont {Das~Sarma}},\ }\bibfield  {title}
  {\enquote {\bibinfo {title} {Machine learning many-body localization: Search
  for the elusive nonergodic metal},}\ }\href {\doibase
  10.1103/PhysRevLett.121.245701} {\bibfield  {journal} {\bibinfo  {journal}
  {Phys. Rev. Lett.}\ }\textbf {\bibinfo {volume} {121}},\ \bibinfo {pages}
  {245701} (\bibinfo {year} {2018})}\BibitemShut {NoStop}%
\bibitem [{\citenamefont {Xu}\ \emph {et~al.}(2019)\citenamefont {Xu},
  \citenamefont {Li}, \citenamefont {Hsu}, \citenamefont {Swingle},\ and\
  \citenamefont {Das~Sarma}}]{butterflyAA}%
  \BibitemOpen
  \bibfield  {author} {\bibinfo {author} {\bibfnamefont {Shenglong}\
  \bibnamefont {Xu}}, \bibinfo {author} {\bibfnamefont {Xiao}\ \bibnamefont
  {Li}}, \bibinfo {author} {\bibfnamefont {Yi-Ting}\ \bibnamefont {Hsu}},
  \bibinfo {author} {\bibfnamefont {Brian}\ \bibnamefont {Swingle}}, \ and\
  \bibinfo {author} {\bibfnamefont {S.}~\bibnamefont {Das~Sarma}},\ }\bibfield
  {title} {\enquote {\bibinfo {title} {Butterfly effect in interacting
  aubry-andre model: Thermalization, slow scrambling, and many-body
  localization},}\ }\href {\doibase 10.1103/PhysRevResearch.1.032039}
  {\bibfield  {journal} {\bibinfo  {journal} {Phys. Rev. Res.}\ }\textbf
  {\bibinfo {volume} {1}},\ \bibinfo {pages} {032039} (\bibinfo {year}
  {2019})}\BibitemShut {NoStop}%
\bibitem [{\citenamefont {Vu}\ \emph {et~al.}(2022)\citenamefont {Vu},
  \citenamefont {Huang}, \citenamefont {Li},\ and\ \citenamefont
  {Das~Sarma}}]{longrangeAA}%
  \BibitemOpen
  \bibfield  {author} {\bibinfo {author} {\bibfnamefont {DinhDuy}\ \bibnamefont
  {Vu}}, \bibinfo {author} {\bibfnamefont {Ke}~\bibnamefont {Huang}}, \bibinfo
  {author} {\bibfnamefont {Xiao}\ \bibnamefont {Li}}, \ and\ \bibinfo {author}
  {\bibfnamefont {S.}~\bibnamefont {Das~Sarma}},\ }\bibfield  {title} {\enquote
  {\bibinfo {title} {Fermionic many-body localization for random and
  quasiperiodic systems in the presence of short- and long-range
  interactions},}\ }\href {\doibase 10.1103/PhysRevLett.128.146601} {\bibfield
  {journal} {\bibinfo  {journal} {Phys. Rev. Lett.}\ }\textbf {\bibinfo
  {volume} {128}},\ \bibinfo {pages} {146601} (\bibinfo {year}
  {2022})}\BibitemShut {NoStop}%
\bibitem [{\citenamefont {Huang}\ \emph
  {et~al.}(2023{\natexlab{a}})\citenamefont {Huang}, \citenamefont {Vu},
  \citenamefont {Li},\ and\ \citenamefont {Das~Sarma}}]{intt1t2}%
  \BibitemOpen
  \bibfield  {author} {\bibinfo {author} {\bibfnamefont {Ke}~\bibnamefont
  {Huang}}, \bibinfo {author} {\bibfnamefont {DinhDuy}\ \bibnamefont {Vu}},
  \bibinfo {author} {\bibfnamefont {Xiao}\ \bibnamefont {Li}}, \ and\ \bibinfo
  {author} {\bibfnamefont {S.}~\bibnamefont {Das~Sarma}},\ }\bibfield  {title}
  {\enquote {\bibinfo {title} {Incommensurate many-body localization in the
  presence of long-range hopping and single-particle mobility edge},}\ }\href
  {\doibase 10.1103/PhysRevB.107.035129} {\bibfield  {journal} {\bibinfo
  {journal} {Phys. Rev. B}\ }\textbf {\bibinfo {volume} {107}},\ \bibinfo
  {pages} {035129} (\bibinfo {year} {2023}{\natexlab{a}})}\BibitemShut
  {NoStop}%
\bibitem [{\citenamefont {Tu}\ \emph {et~al.}(2023{\natexlab{a}})\citenamefont
  {Tu}, \citenamefont {Vu},\ and\ \citenamefont {Das~Sarma}}]{Tu_AA_avalanche}%
  \BibitemOpen
  \bibfield  {author} {\bibinfo {author} {\bibfnamefont {Yi-Ting}\ \bibnamefont
  {Tu}}, \bibinfo {author} {\bibfnamefont {DinhDuy}\ \bibnamefont {Vu}}, \ and\
  \bibinfo {author} {\bibfnamefont {S.}~\bibnamefont {Das~Sarma}},\ }\bibfield
  {title} {\enquote {\bibinfo {title} {Avalanche stability transition in
  interacting quasiperiodic systems},}\ }\href {\doibase
  10.1103/PhysRevB.107.014203} {\bibfield  {journal} {\bibinfo  {journal}
  {Phys. Rev. B}\ }\textbf {\bibinfo {volume} {107}},\ \bibinfo {pages}
  {014203} (\bibinfo {year} {2023}{\natexlab{a}})}\BibitemShut {NoStop}%
\bibitem [{\citenamefont {Schreiber}\ \emph {et~al.}(2015)\citenamefont
  {Schreiber}, \citenamefont {Hodgman}, \citenamefont {Bordia}, \citenamefont
  {L{\"u}schen}, \citenamefont {Fischer}, \citenamefont {Vosk}, \citenamefont
  {Altman}, \citenamefont {Schneider},\ and\ \citenamefont {Bloch}}]{AAexp1}%
  \BibitemOpen
  \bibfield  {author} {\bibinfo {author} {\bibfnamefont {Michael}\ \bibnamefont
  {Schreiber}}, \bibinfo {author} {\bibfnamefont {Sean~S.}\ \bibnamefont
  {Hodgman}}, \bibinfo {author} {\bibfnamefont {Pranjal}\ \bibnamefont
  {Bordia}}, \bibinfo {author} {\bibfnamefont {Henrik~P.}\ \bibnamefont
  {L{\"u}schen}}, \bibinfo {author} {\bibfnamefont {Mark~H.}\ \bibnamefont
  {Fischer}}, \bibinfo {author} {\bibfnamefont {Ronen}\ \bibnamefont {Vosk}},
  \bibinfo {author} {\bibfnamefont {Ehud}\ \bibnamefont {Altman}}, \bibinfo
  {author} {\bibfnamefont {Ulrich}\ \bibnamefont {Schneider}}, \ and\ \bibinfo
  {author} {\bibfnamefont {Immanuel}\ \bibnamefont {Bloch}},\ }\bibfield
  {title} {\enquote {\bibinfo {title} {Observation of many-body localization of
  interacting fermions in a quasirandom optical lattice},}\ }\href {\doibase
  10.1126/science.aaa7432} {\bibfield  {journal} {\bibinfo  {journal}
  {Science}\ }\textbf {\bibinfo {volume} {349}},\ \bibinfo {pages} {842--845}
  (\bibinfo {year} {2015})},\ \Eprint
  {http://arxiv.org/abs/https://www.science.org/doi/pdf/10.1126/science.aaa7432}
  {https://www.science.org/doi/pdf/10.1126/science.aaa7432} \BibitemShut
  {NoStop}%
\bibitem [{\citenamefont {L\"uschen}\ \emph {et~al.}(2017)\citenamefont
  {L\"uschen}, \citenamefont {Bordia}, \citenamefont {Scherg}, \citenamefont
  {Alet}, \citenamefont {Altman}, \citenamefont {Schneider},\ and\
  \citenamefont {Bloch}}]{AAexp2}%
  \BibitemOpen
  \bibfield  {author} {\bibinfo {author} {\bibfnamefont {Henrik~P.}\
  \bibnamefont {L\"uschen}}, \bibinfo {author} {\bibfnamefont {Pranjal}\
  \bibnamefont {Bordia}}, \bibinfo {author} {\bibfnamefont {Sebastian}\
  \bibnamefont {Scherg}}, \bibinfo {author} {\bibfnamefont {Fabien}\
  \bibnamefont {Alet}}, \bibinfo {author} {\bibfnamefont {Ehud}\ \bibnamefont
  {Altman}}, \bibinfo {author} {\bibfnamefont {Ulrich}\ \bibnamefont
  {Schneider}}, \ and\ \bibinfo {author} {\bibfnamefont {Immanuel}\
  \bibnamefont {Bloch}},\ }\bibfield  {title} {\enquote {\bibinfo {title}
  {Observation of slow dynamics near the many-body localization transition in
  one-dimensional quasiperiodic systems},}\ }\href {\doibase
  10.1103/PhysRevLett.119.260401} {\bibfield  {journal} {\bibinfo  {journal}
  {Phys. Rev. Lett.}\ }\textbf {\bibinfo {volume} {119}},\ \bibinfo {pages}
  {260401} (\bibinfo {year} {2017})}\BibitemShut {NoStop}%
\bibitem [{\citenamefont {Kohlert}\ \emph {et~al.}(2019)\citenamefont
  {Kohlert}, \citenamefont {Scherg}, \citenamefont {Li}, \citenamefont
  {L\"uschen}, \citenamefont {Das~Sarma}, \citenamefont {Bloch},\ and\
  \citenamefont {Aidelsburger}}]{GAAexp1}%
  \BibitemOpen
  \bibfield  {author} {\bibinfo {author} {\bibfnamefont {Thomas}\ \bibnamefont
  {Kohlert}}, \bibinfo {author} {\bibfnamefont {Sebastian}\ \bibnamefont
  {Scherg}}, \bibinfo {author} {\bibfnamefont {Xiao}\ \bibnamefont {Li}},
  \bibinfo {author} {\bibfnamefont {Henrik~P.}\ \bibnamefont {L\"uschen}},
  \bibinfo {author} {\bibfnamefont {Sankar}\ \bibnamefont {Das~Sarma}},
  \bibinfo {author} {\bibfnamefont {Immanuel}\ \bibnamefont {Bloch}}, \ and\
  \bibinfo {author} {\bibfnamefont {Monika}\ \bibnamefont {Aidelsburger}},\
  }\bibfield  {title} {\enquote {\bibinfo {title} {Observation of many-body
  localization in a one-dimensional system with a single-particle mobility
  edge},}\ }\href {\doibase 10.1103/PhysRevLett.122.170403} {\bibfield
  {journal} {\bibinfo  {journal} {Phys. Rev. Lett.}\ }\textbf {\bibinfo
  {volume} {122}},\ \bibinfo {pages} {170403} (\bibinfo {year}
  {2019})}\BibitemShut {NoStop}%
\bibitem [{\citenamefont {Ganeshan}\ \emph {et~al.}(2015)\citenamefont
  {Ganeshan}, \citenamefont {Pixley},\ and\ \citenamefont {Das~Sarma}}]{GPD}%
  \BibitemOpen
  \bibfield  {author} {\bibinfo {author} {\bibfnamefont {Sriram}\ \bibnamefont
  {Ganeshan}}, \bibinfo {author} {\bibfnamefont {J.~H.}\ \bibnamefont
  {Pixley}}, \ and\ \bibinfo {author} {\bibfnamefont {S.}~\bibnamefont
  {Das~Sarma}},\ }\bibfield  {title} {\enquote {\bibinfo {title} {Nearest
  neighbor tight binding models with an exact mobility edge in one
  dimension},}\ }\href {\doibase 10.1103/PhysRevLett.114.146601} {\bibfield
  {journal} {\bibinfo  {journal} {Phys. Rev. Lett.}\ }\textbf {\bibinfo
  {volume} {114}},\ \bibinfo {pages} {146601} (\bibinfo {year}
  {2015})}\BibitemShut {NoStop}%
\bibitem [{\citenamefont {{Tu}}\ \emph {et~al.}(2024)\citenamefont {{Tu}},
  \citenamefont {{Long}},\ and\ \citenamefont {{Das
  Sarma}}}]{Tu_NEE_prethermal}%
  \BibitemOpen
  \bibfield  {author} {\bibinfo {author} {\bibfnamefont {Yi-Ting}\ \bibnamefont
  {{Tu}}}, \bibinfo {author} {\bibfnamefont {David~M.}\ \bibnamefont {{Long}}},
  \ and\ \bibinfo {author} {\bibfnamefont {Sankar}\ \bibnamefont {{Das
  Sarma}}},\ }\bibfield  {title} {\enquote {\bibinfo {title} {{Interacting
  quasiperiodic spin chains in the prethermal regime}},}\ }\href {\doibase
  10.48550/arXiv.2405.01622} {\bibfield  {journal} {\bibinfo  {journal} {arXiv
  e-prints}\ ,\ \bibinfo {eid} {arXiv:2405.01622}} (\bibinfo {year} {2024})},\
  \Eprint {http://arxiv.org/abs/2405.01622} {arXiv:2405.01622
  [cond-mat.dis-nn]} \BibitemShut {NoStop}%
\bibitem [{\citenamefont {Tu}\ \emph {et~al.}(2023{\natexlab{b}})\citenamefont
  {Tu}, \citenamefont {Vu},\ and\ \citenamefont
  {Das~Sarma}}]{NEM_avalanche_Tu}%
  \BibitemOpen
  \bibfield  {author} {\bibinfo {author} {\bibfnamefont {Yi-Ting}\ \bibnamefont
  {Tu}}, \bibinfo {author} {\bibfnamefont {DinhDuy}\ \bibnamefont {Vu}}, \ and\
  \bibinfo {author} {\bibfnamefont {Sankar}\ \bibnamefont {Das~Sarma}},\
  }\bibfield  {title} {\enquote {\bibinfo {title} {Localization spectrum of a
  bath-coupled generalized aubry-andr\'e model in the presence of
  interactions},}\ }\href {\doibase 10.1103/PhysRevB.108.064313} {\bibfield
  {journal} {\bibinfo  {journal} {Phys. Rev. B}\ }\textbf {\bibinfo {volume}
  {108}},\ \bibinfo {pages} {064313} (\bibinfo {year}
  {2023}{\natexlab{b}})}\BibitemShut {NoStop}%
\bibitem [{\citenamefont {Lloyd}(1982)}]{lloyd1982least}%
  \BibitemOpen
  \bibfield  {author} {\bibinfo {author} {\bibfnamefont {Stuart}\ \bibnamefont
  {Lloyd}},\ }\bibfield  {title} {\enquote {\bibinfo {title} {Least squares
  quantization in pcm},}\ }\href@noop {} {\bibfield  {journal} {\bibinfo
  {journal} {IEEE transactions on information theory}\ }\textbf {\bibinfo
  {volume} {28}},\ \bibinfo {pages} {129--137} (\bibinfo {year}
  {1982})}\BibitemShut {NoStop}%
\bibitem [{\citenamefont {Bezdek}(1981)}]{bezdek_pattern_1981}%
  \BibitemOpen
  \bibfield  {author} {\bibinfo {author} {\bibfnamefont {James~C.}\
  \bibnamefont {Bezdek}},\ }\href@noop {} {\emph {\bibinfo {title} {Pattern
  recognition with fuzzy objective function algorithms}}},\ \bibinfo {edition}
  {2nd}\ ed.,\ Advanced applications in pattern recognition\ (\bibinfo
  {publisher} {Plenum Press},\ \bibinfo {address} {New York},\ \bibinfo {year}
  {1981})\BibitemShut {NoStop}%
\bibitem [{Note1()}]{Note1}%
  \BibitemOpen
  \bibinfo {note} {Eigenstates of the top band edge of the spectrum resemble
  those from the bottom band edge of the spectrum.}\BibitemShut {Stop}%
\bibitem [{\citenamefont {Pearson}(1901)}]{pearson1901liii}%
  \BibitemOpen
  \bibfield  {author} {\bibinfo {author} {\bibfnamefont {Karl}\ \bibnamefont
  {Pearson}},\ }\bibfield  {title} {\enquote {\bibinfo {title} {Liii. on lines
  and planes of closest fit to systems of points in space},}\ }\href@noop {}
  {\bibfield  {journal} {\bibinfo  {journal} {The London, Edinburgh, and Dublin
  philosophical magazine and journal of science}\ }\textbf {\bibinfo {volume}
  {2}},\ \bibinfo {pages} {559--572} (\bibinfo {year} {1901})}\BibitemShut
  {NoStop}%
\bibitem [{\citenamefont {Tibshirani}\ \emph {et~al.}(2001)\citenamefont
  {Tibshirani}, \citenamefont {Walther},\ and\ \citenamefont
  {Hastie}}]{tibshirani2001estimating}%
  \BibitemOpen
  \bibfield  {author} {\bibinfo {author} {\bibfnamefont {Robert}\ \bibnamefont
  {Tibshirani}}, \bibinfo {author} {\bibfnamefont {Guenther}\ \bibnamefont
  {Walther}}, \ and\ \bibinfo {author} {\bibfnamefont {Trevor}\ \bibnamefont
  {Hastie}},\ }\bibfield  {title} {\enquote {\bibinfo {title} {Estimating the
  number of clusters in a data set via the gap statistic},}\ }\href@noop {}
  {\bibfield  {journal} {\bibinfo  {journal} {Journal of the Royal Statistical
  Society: Series B (Statistical Methodology)}\ }\textbf {\bibinfo {volume}
  {63}},\ \bibinfo {pages} {411--423} (\bibinfo {year} {2001})}\BibitemShut
  {NoStop}%
\bibitem [{\citenamefont {Serbyn}\ \emph {et~al.}(2016)\citenamefont {Serbyn},
  \citenamefont {Michailidis}, \citenamefont {Abanin},\ and\ \citenamefont
  {Papi\ifmmode~\acute{c}\else \'{c}\fi{}}}]{ES_MBL}%
  \BibitemOpen
  \bibfield  {author} {\bibinfo {author} {\bibfnamefont {Maksym}\ \bibnamefont
  {Serbyn}}, \bibinfo {author} {\bibfnamefont {Alexios~A.}\ \bibnamefont
  {Michailidis}}, \bibinfo {author} {\bibfnamefont {Dmitry~A.}\ \bibnamefont
  {Abanin}}, \ and\ \bibinfo {author} {\bibfnamefont {Z.}~\bibnamefont
  {Papi\ifmmode~\acute{c}\else \'{c}\fi{}}},\ }\bibfield  {title} {\enquote
  {\bibinfo {title} {Power-law entanglement spectrum in many-body localized
  phases},}\ }\href {\doibase 10.1103/PhysRevLett.117.160601} {\bibfield
  {journal} {\bibinfo  {journal} {Phys. Rev. Lett.}\ }\textbf {\bibinfo
  {volume} {117}},\ \bibinfo {pages} {160601} (\bibinfo {year}
  {2016})}\BibitemShut {NoStop}%
\bibitem [{\citenamefont {Yang}\ \emph {et~al.}(2015)\citenamefont {Yang},
  \citenamefont {Chamon}, \citenamefont {Hamma},\ and\ \citenamefont
  {Mucciolo}}]{ES_ETH}%
  \BibitemOpen
  \bibfield  {author} {\bibinfo {author} {\bibfnamefont {Zhi-Cheng}\
  \bibnamefont {Yang}}, \bibinfo {author} {\bibfnamefont {Claudio}\
  \bibnamefont {Chamon}}, \bibinfo {author} {\bibfnamefont {Alioscia}\
  \bibnamefont {Hamma}}, \ and\ \bibinfo {author} {\bibfnamefont {Eduardo~R.}\
  \bibnamefont {Mucciolo}},\ }\bibfield  {title} {\enquote {\bibinfo {title}
  {Two-component structure in the entanglement spectrum of highly excited
  states},}\ }\href {\doibase 10.1103/PhysRevLett.115.267206} {\bibfield
  {journal} {\bibinfo  {journal} {Phys. Rev. Lett.}\ }\textbf {\bibinfo
  {volume} {115}},\ \bibinfo {pages} {267206} (\bibinfo {year}
  {2015})}\BibitemShut {NoStop}%
\bibitem [{\citenamefont {Huang}\ \emph
  {et~al.}(2023{\natexlab{b}})\citenamefont {Huang}, \citenamefont {Vu},
  \citenamefont {Li},\ and\ \citenamefont {Das~Sarma}}]{t1t2}%
  \BibitemOpen
  \bibfield  {author} {\bibinfo {author} {\bibfnamefont {Ke}~\bibnamefont
  {Huang}}, \bibinfo {author} {\bibfnamefont {DinhDuy}\ \bibnamefont {Vu}},
  \bibinfo {author} {\bibfnamefont {Xiao}\ \bibnamefont {Li}}, \ and\ \bibinfo
  {author} {\bibfnamefont {S.}~\bibnamefont {Das~Sarma}},\ }\bibfield  {title}
  {\enquote {\bibinfo {title} {Incommensurate many-body localization in the
  presence of long-range hopping and single-particle mobility edge},}\ }\href
  {\doibase 10.1103/PhysRevB.107.035129} {\bibfield  {journal} {\bibinfo
  {journal} {Phys. Rev. B}\ }\textbf {\bibinfo {volume} {107}},\ \bibinfo
  {pages} {035129} (\bibinfo {year} {2023}{\natexlab{b}})}\BibitemShut
  {NoStop}%
\end{thebibliography}
%

\end{document}